\documentclass[twocolumn,10pt]{article}

\usepackage[a4paper,margin=0.7in]{geometry}
\usepackage[english]{babel}
\usepackage{float}
\usepackage[dvipsnames]{xcolor}
\usepackage{graphicx}
\usepackage{amsmath, amsthm, amssymb}
\usepackage{amsfonts}
\usepackage{romannum}
\usepackage{tikz}
\usepackage{titlesec}
\usepackage{abstract}
\usepackage[numbers,sort&compress]{natbib}
\usepackage{caption}
\usepackage{subcaption}
\usepackage{mathrsfs} 
\usepackage{upgreek}
\usepackage[T1]{fontenc}
\usepackage{lmodern}

\usepackage[
    colorlinks=true,
    linkcolor=red,
    filecolor=blue,
    urlcolor=blue,
    citecolor=blue,
    breaklinks=true
]{hyperref}

\newcommand{\orcid}[1]{%
  \,\href{https://orcid.org/#1}{\raisebox{-0.4pt}{\includegraphics[width=8pt]{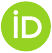}}}%
}

\newcommand{\ii}{\mathrm{i}}
\newcommand{\AC}{\mathrm{AC}}
\newenvironment{onecoltextblock}[1][]
{%
  \begin{center}
  \begin{minipage}{0.92\textwidth}
  \if\relax\detokenize{#1}\relax
  \else
    {\centering\large\bfseries #1\par\vspace{0.6em}}
  \fi
  \small
}
{%
  \end{minipage}
  \end{center}
}

\renewcommand{\thesection}{\Roman{section}}
\renewcommand{\thesubsection}{\Roman{section}.\Alph{subsection}}

\titleformat{\section}
  {\large\bfseries\fontsize{10}{12}\selectfont\MakeUppercase}
  {\thesection.}{0.5em}{}

\titleformat{\subsection}
  {\normalsize\bfseries}
  {\thesubsection.}{0.5em}{}

\title{%
\vspace{-1.5em}
\textbf{\fontsize{16pt}{16pt}\selectfont\bfseries Scanless quantum Fourier-transform mid-infrared spectroscopy for rapid high-sensitivity hyperspectral mapping}
\vspace{-0.5em}
}

\author{%
\parbox{0.96\textwidth}{\centering
\normalsize
Paul Gattinger\orcid{0000-0002-0120-9100}$^{1,*}$\quad
Bettina Heise\orcid{0000-0001-9978-0761}$^{1}$\quad
Andreas W. Schell\orcid{0000-0003-0849-9558}$^{2}$\\[0.25em]
Kristina Duswald\orcid{0000-0002-0445-6967}$^{1}$\quad
Markus Brandstetter\orcid{0000-0002-8679-8097}$^{1}$\quad
Ivan Zorin\orcid{0000-0002-2089-5716}$^{1,*}$\\[0.8em]
\footnotesize\itshape
$^{1}$Research Center for Non-Destructive Testing, Science Park 2, Altenberger Str. 69, 4040 Linz, Austria\\
$^{2}$Division of Light-Matter-Interaction, Johannes Kepler University, Altenberger Str. 69, 4040 Linz, Austria\\[0.6em]
\footnotesize\normalfont
$^{*}$Corresponding authors: \href{mailto:paul.gattinger@recendt.at}{paul.gattinger@recendt.at} and \href{mailto:ivan.zorin@recendt.at}{ivan.zorin@recendt.at}\\[0.6em]
\footnotesize
\today
}
}
\date{}

\begin{document}
\twocolumn[
\maketitle
\vspace{-2em}

\begin{onecolabstract} 
Fourier-transform infrared (FTIR) spectroscopy is a well-established technique for qualitative and quantitative chemical analysis. Classical FTIR systems rely, however, on direct mid-infrared (mid-IR) scan-based time-domain measurements of coherence functions; thus, the signal-to-noise ratio and measurement speed are constrained by design.
In this paper, we demonstrate a scanless quantum FTIR (sQFTIR) technique that exploits principles of metrology with entangled photons to circumvent the limitations inherent to classical FTIR systems. 
The approach exploits the interferometric nature of the sensing paradigm and relies on frequency-domain measurements performed with a static, low-gain nonlinear interferometer. 
A robust reconstruction algorithm is used to retrieve time-domain signals and reconstruct respective mid-infrared (mid-IR) spectra (3000~cm\textsuperscript{-1} to 2380~cm\textsuperscript{-1}) from near-IR measurements (approx. 780~nm to 820~nm).
The suggested sQFTIR protocol eliminates the need for optical delay scanning and leverages inherent mapping between the related domains. In the theoretical section, we evaluate the intrinsic signal-to-noise advantage of the proposed method over conventional scan-based time-domain measurements; a difference of 26.8 dB (factor of 21.8) is demonstrated. Building on the enhanced sensitivity of the scheme, we demonstrate rapid sQFTIR-based hyperspectral imaging with a spatial resolution of 12.3~\textmu m and a spectral resolution down to 8~cm\textsuperscript{-1}. Hyperspectral mapping of human colon tissue, microplastics, and multilayer polymer samples composed of polypropylene and ethylene vinyl alcohol yield high-quality single-pixel spectra with acquisition times down to 10~ms. 
\end{onecolabstract}
\vspace{0.5em}
]\saythanks

\section{Introduction}
 
Fourier-transform infrared spectroscopy (FTIR) is a well-established analytical technique that employs principles of low-coherence interferometry to access fundamental molecular vibrations for non-destructive, label-free qualitative and quantitative chemical analysis~\cite{griffiths2007fourier, chalmers_handbook_2001}. Variations of FTIR are well-adopted across fundamental as well as applied fields, ranging from studies of molecular structures~\cite{https://doi.org/10.1002/bip.360250307,surewicz_determination_1993} and biomedical applications~\cite{Magalhes26112021,Lopes21102018,Movasaghi01022008,xie_analysing_2024} (including imaging~\cite{fernandez_infrared_2005} and diagnostics~\cite{sitnikova_breast_2020}) to non-destructive evaluation~\cite{liu_recent_2022}, gas sensing~\cite{https://doi.org/10.1002/lpor.202100556}, and process monitoring~\cite{rolinger_critical_2020,koch_comparison_2016} in industrial facilities. However, the evolution of classical FTIR experienced over more than half a century has become increasingly constrained by the persistent challenges associated with classical mid-IR sources and detectors~\cite{Rogalski2010}. This limitation has stimulated the development of laser-driven~\cite{curl_quantum_2010, schwaighofer_quantum_2017,Zorin:22}, up-conversion~\cite{Tidemand-Lichtenberg:16, Barh:17}, and most recently, quantum-based methods\textemdash a particularly promising emerging technique is sensing with undetected photons.

Sensing with undetected photons is a quantum-optical measurement approach that exploits photon entanglement and bi-photon interference to separate the spectral domains of probing and detection~\cite{Lemos2014, BarretoLemos:22}. In recent years, many compelling implementations of sensing with undetected mid-IR photons using visible or near-IR detection have been demonstrated, such as frequency-domain spectroscopy~\cite{Kalashnikov2016,Kaufmann2022}, optical coherence tomography (OCT)~\cite{Paterova2018,Vanselow2020,Zotti:25}, quantum FTIR (QFTIR) spectroscopy~\cite{Lindner:21,Mukai2022,Lindner:22, Lindner2023,Kaur2024,https://doi.org/10.1002/qute.202300299, Gattinger2025}, imaging~\cite{Lemos2014,Viswanathan2021,doi:10.1126/sciadv.abd0264,Kviatkovsky:22,PhysRevApplied.19.054019,LeonTorres2025} and hyperspectral imaging~\cite{Placke2026}. 
In essence, metrology with undetected photons relies on quantum low-coherence (white-light) interferometry~\cite{10.1063/5.0004873,Zou1991,Chekhova2016} and is fundamentally rooted in the principle of indistinguishability. The method employs a quasi-monochromatic pump laser and two subsequent parametric amplifiers (e.g. $\chi^{(2)}$ crystals) operated in the non-degenerate regime. Hence, broadband signal–idler photon pairs are generated by spontaneous parametric down-conversion (SPDC) in two indistinguishable parametric interactions. The resulting interference pattern recorded in the spectral domain of the signal photons is determined by the phase and transmission experienced by the corresponding idler photons, thereby encoding mid-IR sample information into the signal wavelengths. This interference can be measured either in the frequency domain using a dispersive spectrometer or in the time domain by scanning the relative optical delay.

Frequency-domain sensing with undetected photons uniquely combines spectral-domain acquisition with intrinsically interferometric readout, enabling spectral and spatial (optical delay) information encoded in the interferometric fringes to be acquired simultaneously, analogous to spectral-domain and spectroscopic optical coherence tomography (OCT)~\cite{drexler_optical_2015,Vanselow2020}. This stands in contrast to classical mid-IR spectroscopy, where dispersive spectrometers and interferometric FTIR systems generally represent redundant rather than complementary approaches. 
In this work, the distinctive multiplexing capabilities of frequency-domain sampling are exploited through a tailored reconstruction protocol to eliminate mechanical scanning and introduce scanless quantum FTIR (sQFTIR).
The sQFTIR acquisition scheme has similarities to Hilbert-transform-based envelope-extraction approaches~\cite{Kaufmann2022, Sherwani:26}.
However, the proposed method aims to retrieve time-domain signals akin to those obtained in classical FTIR and to reconstruct mid-IR spectra, thereby facilitating high-speed, robust, and high-sensitivity sampling.
Furthermore, sQFTIR frequency-domain acquisition in the shot-noise-dominated regime features an intrinsic sensitivity advantage unattainable in classical direct mid-IR spectroscopy. Kaufmann et al.~\cite{Kaufmann2022} argued that Fellgett's advantage, which is crucial in classical FTIR spectroscopy~\cite{Saptari2003} when signal-independent detector noise dominates, is not applicable (or strongly reduced) in schemes based on sensing with undetected photons due to the shot-noise-limited performance of silicon-based arrays and detectors. In this paper, we extend the analysis further and demonstrate that sensing in the Fourier domain, in fact, provides an intrinsic sensitivity advantage over time-domain approaches for comparable measurement times due to a multiplex advantage associated with multipixel detection and coherent averaging. The advantage, well-known in classical optical coherence tomography (OCT), is enabled by the inherent interferometric configuration coupled to detection of spectral interferograms that provides coherent addition of all depth and spectral components at once.\\
Here, we address limitations of classical scan-based FTIR as well as QFTIR and propose an efficient sQFTIR modality.
The proposed scanless nonlinear interferometric technique relies on two fundamental mathematical principles: (i) the Wiener–Khinchin theorem (including its complex generalization), which relates the time (spatial) and Fourier (spectral) domains and manifests that measurements (coherence functions and spectra) can be performed in either of them; and (ii) Parseval’s theorem, which ensures the conservation of energy and noise power across these domains~\cite{Mandel_Wolf_1995,bracewell_fourier_2000}.
A tailored reconstruction protocol based on sequential Fourier transforms to retrieve time-domain central bursts and mid-IR spectra is introduced. Furthermore, we provide a comprehensive analysis and quantification of the sensitivity advantage and present relevant experimental confirmation. 

Due to the high sensitivity, single-mode operation, and robustness of the method the experimental demonstration is extended to microscopic hyperspectral sQFTIR-based mapping of application-relevant samples such as multilayer polymer foils, microplastics, and human tissue. 

\section{Scanless quantum FTIR}

\subsection{Principles and reconstruction protocol}

QFTIR spectroscopy\textemdash the quantum-optical analogue to classical FTIR spectroscopy\textemdash works without the use of native IR sources and detectors. Instead, visible or near-IR light sources and detectors are employed. This is enabled by the underlying quantum phenomenon of bi-photon interference~\cite{Zou1991,Chekhova2016}. 
The core unit of the sQFTIR modality is a nonlinear SU(1,1) interferometer with a Michelson topology, schematically shown in Fig.~\ref{fig:su11-scheme} (linear configuration, for experimental implementation see Section~\ref{sec:exp}). It should be noted that, in the following, we focus only on the low-gain regime.
In this arrangement, pump photons, $\omega_p$, are converted with a low probability ($\mathrm{p}\approx 10^{-9}$) into energy-correlated signal $\omega_s$ and idler $\omega_i$ photons in the forward or backward pass through a nonlinear crystal (NL), so that $\omega_p=\omega_s+\omega_i$.
\begin{figure}[ht]
\centering
\includegraphics[width=0.85\linewidth]{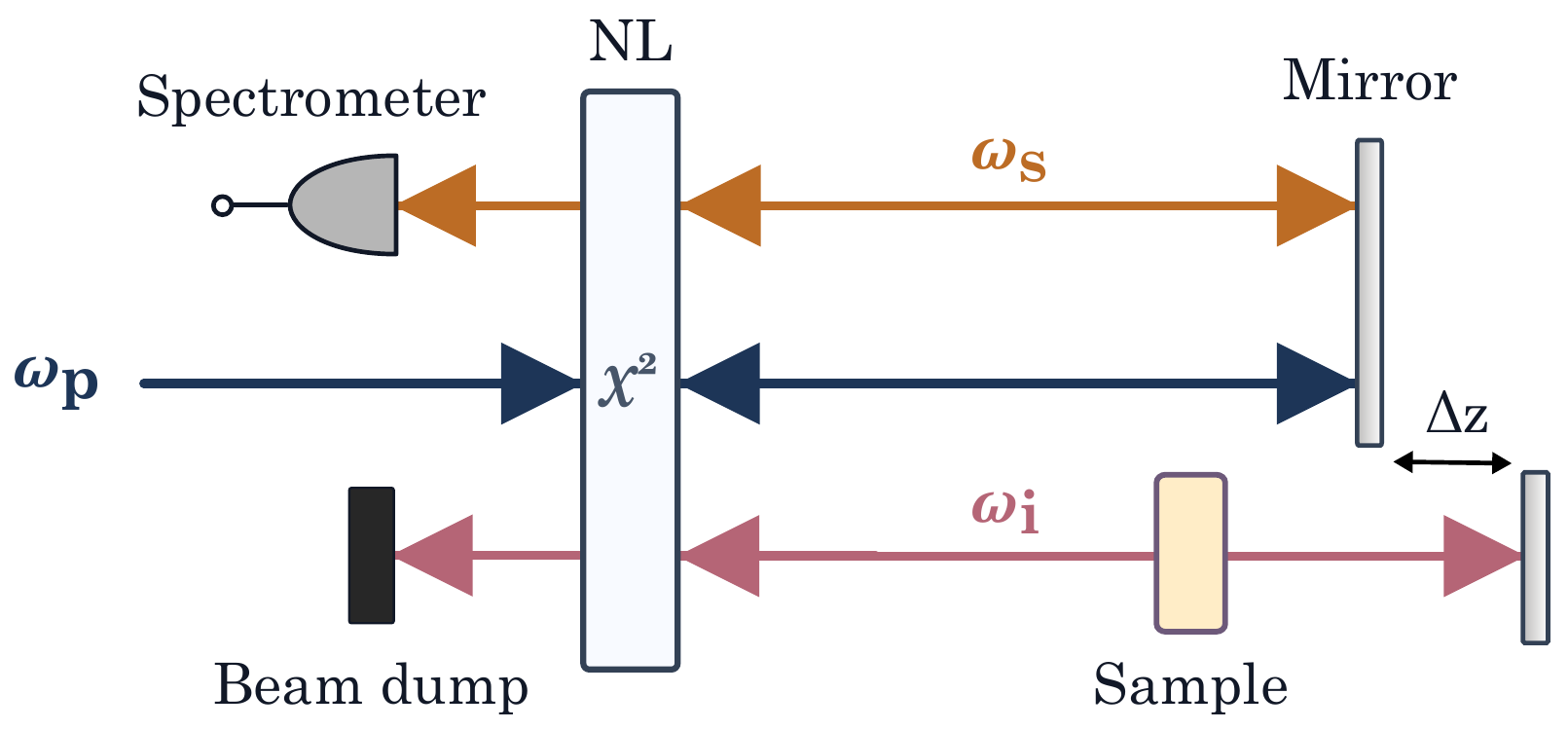}
\caption{Linear schematic of sQFTIR spectroscopy. Pump photons $\omega_p$ are converted (with a low probability) to broadband entangled signal, $\omega_s$ and idler, $\omega_i$ photons in the forward or backward path through a nonlinear crystal (NL). Due to the indistinguishability of the photon paths, interference between $\omega_s$ photons generated in forward and backward propagation can be observed at the spectrometer. This spectral interference pattern is defined by the path difference $\Delta z$. A sample inserted in the $\omega_i$ path disturbs the interference pattern (phase shift or loss of interference, i.e. absorption)} \label{fig:su11-scheme}
\end{figure}

The $\omega_i$ photon path is offset by a relative optical delay $\Delta z$ between the first and second pass through the NL; this is a crucial step for the sQFTIR method. Due to the indistinguishability of the photon paths (which-path or which-source information is lost after the second pass), a FD interference pattern (spectral interferogram) can be observed by coupling a spectrometer to the $\omega_s$ photon output of the nonlinear interferometer. This spectral interferogram encodes all depth and spectral information simultaneously. Phase and absorption information of a sample inserted in the $\omega_i$ photon path is imprinted in the $\omega_s$ spectral interferogram. This type of interference pattern is unique to this kind of quantum-based interferometry. It strongly resembles classical FD-OCT spectra; however, if one blocks the $\omega_i$ photons, the number of detected signal photons remains unchanged, since SPDC is not a seeded process. In classical FD-OCT, blocking the sample arm would lead to a significant drop in the signal, depending on the ratio of the beam splitter.
\begin{figure*}[hbt]
\centering
\includegraphics[width=.95\textwidth]{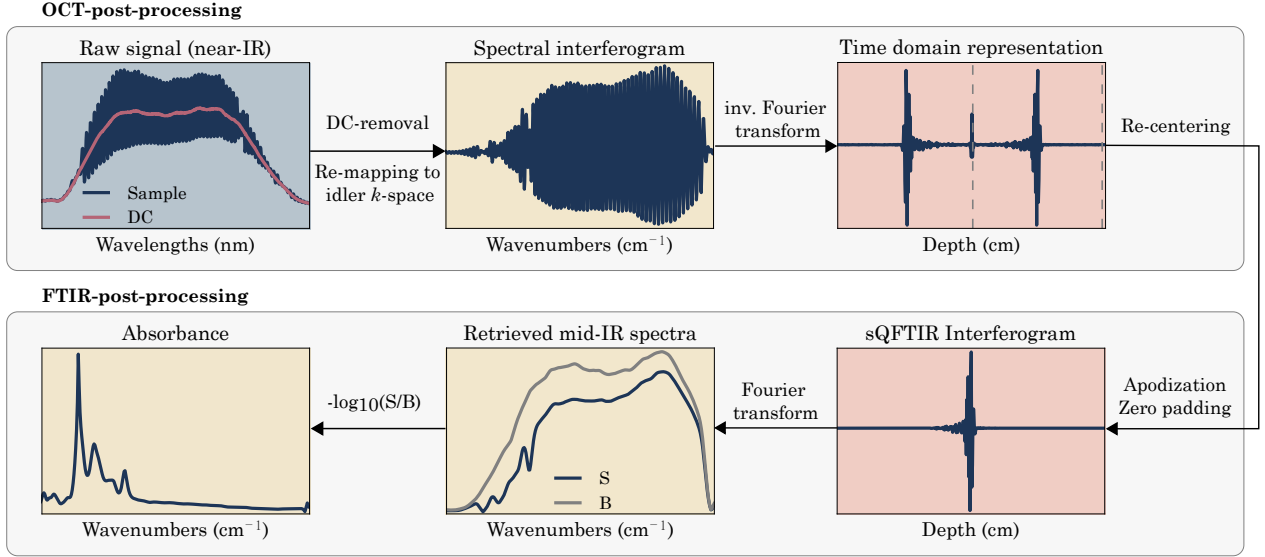}
\caption{Processing steps of sQFTIR spectroscopy. A raw spectral interferogram is DC corrected, remapped to $k$-space; an inverse Fourier transform is then applied (OCT post-processing). The coherence burst of a single reflector (i.e., the sample) is isolated, centered, apodized, zero-padded and Fourier transformed. The demodulated sample and background spectra are further processed according to standard IR spectroscopy protocols in order to receive absorbance spectra.} \label{fig:flow-diagram}
\end{figure*}
The first demonstrations of IR spectroscopy~\cite{Kalashnikov2016,Kaufmann2022} and OCT~\cite{Vanselow2020} with undetected photons relied on measurements in the FD. For OCT with undetected photons, the post-processing is analogous to the classical counterpart. However, for quantum-based IR spectroscopy different approaches, e.g. Hilbert-transform-based~\cite{Kaufmann2022,Sherwani:26}, have to be used to extract the absorption magnitude from the interference fringes. 

In contrast, QFTIR spectroscopy is a time-domain-based (TD-based) method and either the idler~\cite{Lindner2023,Mukai2022}, or combined signal and pump~\cite{Gattinger2025} phase is scanned. The TD interferogram (coherence function, or so-called central burst) is measured with a single-pixel detector in the $\omega_s$ spectral domain. If a sample is placed in the idler photon path, its absorption and phase information are imprinted in the ($\omega_s$) center burst. 

In the following, we introduce a method, based on well-established, computationally efficient and robust reconstruction algorithms, to retain the spectral information from raw FD spectra by using two subsequent Fourier transforms. A general workflow of the sQFTIR approach is outlined in Fig.~\ref{fig:flow-diagram}; it adapts and combines the general post-processing steps of classical OCT and FTIR spectroscopy.  

The raw FD interferogram and a DC measurement (mean SPDC profile after second pass), for which the idler photons are blocked, are recorded. The DC component is subtracted from the FD spectrum, resulting in a zero-centered spectral interferogram. The resulting spectral interferogram is remapped to wavenumbers (k-space); this is an OCT post-processing step to ensure linearity within the target k-space~\cite{Wojtkowski2002}. After an inverse Fourier transform, re-centering to a single coherence burst of the symmetric TD representation is performed, thus leading to the FTIR-post-processing steps. The isolated and centered signal from the OCT-post-processing steps is equivalent to a TD center burst typical in FTIR and QFTIR spectroscopy. This center burst is apodized, zero-padded and then Fourier transformed. The magnitude of this Fourier transform represents the mid-IR (idler-emission) spectrum. After measuring a spectrum of the empty interferometer (B) and one with a sample inserted (S), absorbance can be computed according to the convention in infrared spectroscopy $A = -\mathrm{log_{10}}(\mathrm{S}/\mathrm{B})$. Similarly, dispersion measurements can be performed. This is essentially QFTIR spectroscopy without scanning of the relative phases. The spectral resolution $\delta \tilde{\nu}_i$ of this method depends on the selected group delay, $\Delta z_w$, offset from the zero-delay plane, which translates to the fringe speed in the FD measurement ($k$-space).

This demodulation procedure can be described in terms of the Fourier-shift theorem. The initial modulation in the spectral interferogram can be attributed to a linear phase factor due to a shift of the coherence bursts in the TD. By selecting and centering one of the bursts prior to the second Fourier transform, the initial phase factor is compensated and thus the spectrum is demodulated after the transform. The details and mathematical description of the flow diagram shown in Fig.~\ref{fig:flow-diagram} can be found in Supplementary 1. A Python implementation of the post-processing protocol, containing scripts for processing as well as raw data, is available in a public repository~\cite{sqFTIR-git}.

It should be emphasized that the method has similarities to so-called spectroscopic OCT. The spectral resolution (sQFTIR, $\Delta k$) and the depth resolution (axial resolution in OCT, $\Delta z$) are governed by the following relation
\begin{equation}
\Delta k \ \Delta z \ge \mathrm{const.}
\end{equation}
This formula describes the balance between axial and spectral resolution that has to be considered thoroughly in spectroscopic OCT. However, in the case of sQFTIR spectroscopy, the axial (spatial) resolution ($\Delta z$) is sacrificed in order to achieve high spectral resolution ($\Delta k$).

\subsection{Sensitivity advantage of Fourier-domain detection}\label{sec:sensadv}
In classical mid-IR spectroscopy, FTIR instruments are regarded as the gold standard for spectral measurements because they offer two principal sensitivity advantages over dispersive systems based on gratings or prisms: the Jacquinot throughput advantage and the Fellgett multiplex advantage~\cite{Saptari2003}. These advantages mainly stem from the fact that IR detectors have low quantum efficiency and high detector noise, including thermal noise, 1/f noise, and other noise sources~\cite{Rogalski2010}. However, it is well known from the OCT literature that in the visible and near-IR spectral domains the possibility of shot-noise-limited measurements turns the Fellgett advantage into a disadvantage, meaning that TD measurements in these spectral domains are less sensitive than FD measurements under a fixed measurement time. In the following section, we adapt a framework from the OCT literature~\cite{Choma:03,Leitgeb2003, DeBoer2017} and apply it to nonlinear interferometry to describe the advantage of FD (dispersive) measurements over TD measurements in the shot-noise-limited regime.  

Parseval's theorem states that the total power measured in the TD is conserved under Fourier transformation to the FD and vice versa:
\begin{equation}
P_{\mathrm{tot}} = \int_{-\infty}^{\infty}|s(z)|^2 dz = \int_{-\infty}^{\infty}|S(\nu)|^2 d\nu.
\end{equation}
Here, $s(z)$ and $S(\nu)$ are the signal representations in the TD ($z$, distance/optical delay space) and the FD ($\nu$, optical frequency), respectively. Parseval’s theorem ensures that a unitary Fourier transformation preserves the total signal energy and noise power; hence, a sensitivity advantage observed in one domain persists under transformation to the conjugate domain.

The expression for the interferometric signal from a single reflector, measured by a photodetector in the TD, can be adapted for nonlinear interferometry using the formalism from~\cite{Sorin1992} (only signal photons are measured) as follows:
\begin{equation}
I(z) = \frac{2\eta e P_s}{E_\nu} \left [1 +  \upnu \cos{\Delta \phi(z)} \right] \quad [e/s],
\end{equation}
where $\eta$ is the efficiency of the detector, $\upnu$ is the interference visibility, $e$ is the electron charge, $E_{\nu}$ is the photon energy, $P_s$ is the signal power generated in one optical parametric interaction, and $\Delta \phi(z)$ is the phase delay of pump, signal, and idler photons (optical delay). The physical dimensions of \(I(z)\) indicate that it represents a photocurrent, as a photodiode is used for detection. $P_s$ represents the probability $|f(\omega_s)|^2$ of photon emission into a certain mode $\omega_s$.
The signal for the same reflector in the FD can be represented as
\begin{equation}
I(\nu)
=
\frac{2\eta e \tau P_s}{N E_\nu}
\left[
1+\upnu\cos{\Delta\phi(\nu)}
\right]
\quad
\left[\mathrm{e}\right],
\end{equation}
where $N$ is the number of spectral channels and $\tau$ is the integration time per channel. It has the physical dimensions of a charge, since detection is performed using CCD or CMOS detector technologies. 

The signal power in the signal-processing sense (with the noninterferometric DC term removed) in the TD can be written as~\cite{DeBoer2017}
\begin{equation}
S_{\mathrm{TD}} = \bigl \langle I(z)^2 \bigr \rangle = 2 \biggl( \frac{\upnu \eta e}{E_\nu} \biggr)^2
P_s^2 \quad [(e/s)^2],
\end{equation}
and the corresponding FD representation as 
\begin{equation}
S_{\mathrm{FD}} = \bigl \langle I(\nu)^2 \bigr \rangle = 2 \biggl( \frac{\upnu \eta e \tau}{E_\nu} \biggr)^2
\frac{P_s^2}{N^2} \quad [e^2].
\end{equation}
The shot-noise power in each domain can be expressed as the respective variances
\begin{equation}
\sigma^2_{\mathrm{shot,TD}} = \frac{2B \eta e^2}{E_\nu} 2P_{s}, \quad \sigma^2_{\mathrm{{shot,FD}}} = \frac{\eta e^2 \tau}{N E_\nu} 2P_{s},
\end{equation}
where B is the detector bandwidth. $\sigma^2_{\mathrm{shot,TD}}$ is expressed as an electric current variance in A$^2$, whereas the $\sigma^2_{\mathrm{{shot,FD}}}$ can be interpreted as a charge variance in C$^2$. In contrast to classical OCT, the total number of detected signal photons is constant and corresponds to the double pass through the crystal, i.e., $2P_s$.
The shot noise is conserved, meaning that the noise power per detector element of the spectrometer is equal to the noise power per virtual detector element after the Fourier transform
\begin{equation}
\sigma^2_{\mathrm{shot}}(z) = \sigma^2_{\mathrm{shot}}(\nu).
\end{equation}
This is a valid assumption due to the conservation of total power under a Fourier transform, as described by Parseval's theorem, and the fact that shot noise is white and follows Poisson statistics. 

It is conventional in OCT to express the signal-to-noise ratio (SNR) in terms of a power ratio $(i^2/\sigma^2)$, where $i$ represents the photocurrent or charge and $\sigma $ is the standard deviation of the noise. Therefore, the total noise can be approximated by simple addition of individually contributing noise power sources
\begin{equation}
\sigma^2 = \sigma^2_{r+d} + \sigma^2_{\mathrm{shot}} +\sigma^2_{\mathrm{{RIN}}}. 
\end{equation}
Here, $\sigma^2_{r+d}$ denotes the combined readout and dark noise, and $\sigma^2_{\mathrm{{RIN}}}$ is the relative intensity noise, which usually plays a role at high source powers. In the ideal case of infinite bandwidth and a perfect reflector, the FD signal translates to the time (spatial) domain upon Fourier transform in the following way: it is concentrated in two pixels, one at the positive depth position $z$ and one at the negative depth position $-z$
\begin{equation}
S_{\mathrm{FD}}(z)=S_{\mathrm{FD}}(-z) = \frac{N S_{\mathrm{FD}}(\nu)}{2} =  \biggl( \frac{\upnu \eta e \tau}{E_\nu} \biggr)^2 \frac{P_s^2}{N}.
\end{equation}
Thus, writing the SNR in the FD as a power ratio gives

\begin{equation}
\mathrm{SNR}_{\mathrm{FD}}
=
\frac{
    \displaystyle
    \left(
        \dfrac{\upnu \eta e \tau}{E_\nu}
    \right)^2
    \dfrac{P_s^2}{N}
}{
    \displaystyle
    \sigma_{r+d}^2
    +
    \dfrac{\eta e^2 \tau}{N E_\nu} 2P_s
    +
    \sigma_{\mathrm{RIN}}^2
}.\label{eq:fullSNR}
\end{equation}
In the case of shot-noise-limited detection, for which $\sigma^2_{\mathrm{shot}} \gg \sigma^2_{r+d} + \sigma^2_{\mathrm{{RIN}}}$, the resulting SNR for FD and TD measurements can be expressed in terms of power ($i^2/\sigma^2$, OCT standard) as
\begin{equation}
\mathrm{SNR}_{\mathrm{FD}} = \frac{\upnu^2 \eta P_s \tau}{2E_\nu}, \quad \mathrm{SNR}_{\mathrm{TD}} = \frac{\upnu ^2\eta P_s}{2E_\nu B},
\label{eq:SNR}
\end{equation}
and in terms of amplitude ($i/\sigma$, spectroscopy standard) as
\begin{equation}
\mathrm{SNR}_{\mathrm{FD}}^{[i/\sigma]} = \upnu \sqrt{\frac{\eta P_s \tau}{2E_\nu}}, \quad \mathrm{SNR}_{\mathrm{TD}}^{[i/\sigma]} = \upnu \sqrt{\frac{\eta P_s}{2E_\nu B}}.
\label{eq:SNR_spectroscopy}
\end{equation}

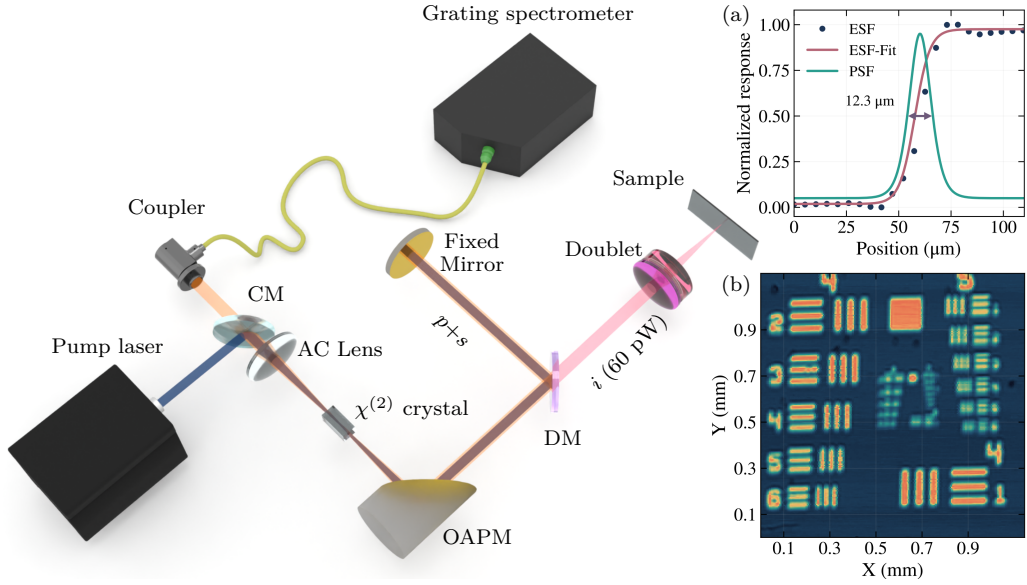
\begin{figure*}[htb!]
\centering
\begin{tikzpicture}
  \node[anchor=south west,inner sep=0] (image) at (0,0)
    {\includegraphics[width=0.8\textwidth]{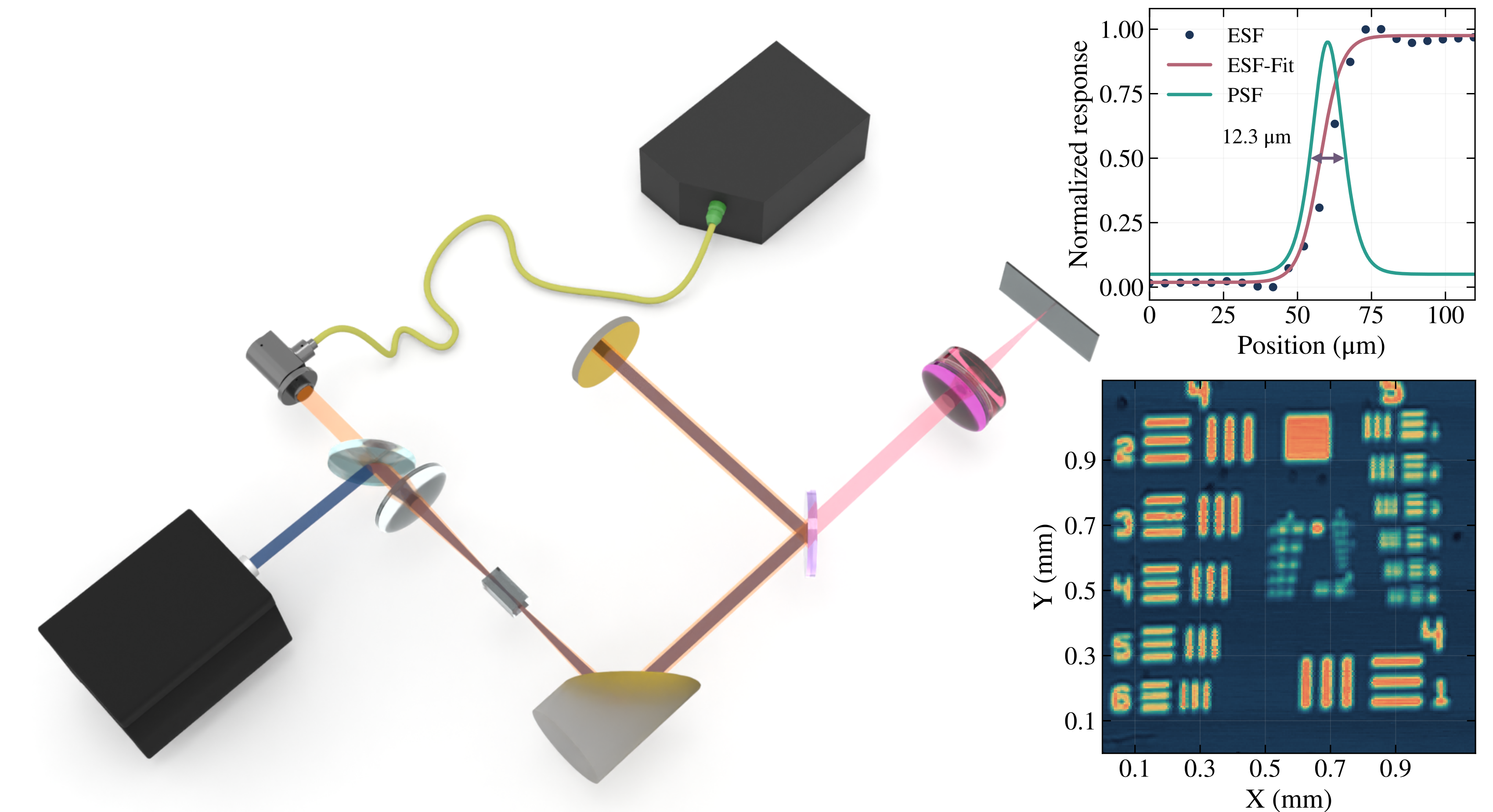}};
  \begin{scope}[x={(image.south east)},y={(image.north west)}]
    \draw (0.12,0.428-0.025) node[]{\color{black}\footnotesize Pump laser};
    \draw (0.176,0.67-0.025) node[]{\color{black}\footnotesize Coupler};
    \draw (0.270,0.522-0.025) node[]{\color{black}\footnotesize CM};
    \draw (0.34,0.426-0.025) node[]{\color{black}\footnotesize AC Lens};
    \draw (0.408,0.324-0.025) node[]{\color{black}\footnotesize $\chi^{(2)}$ crystal};
    \draw (0.47,0.1-0.025) node[]{\color{black}\footnotesize OAPM};
    \draw (0.551,0.271-0.025) node[]{\color{black}\footnotesize DM};
    \draw (0.465,0.614-0.025) node[]{\color{black}\footnotesize Fixed};
    \draw (0.465,0.614-0.025-0.04) node[]{\color{black}\footnotesize Mirror};
    \draw (0.59,0.6-0.025) node[]{\color{black}\footnotesize Doublet};
    \draw (0.632,0.72-0.025) node[]{\color{black}\footnotesize Sample};
    \draw (0.614,0.394) node[rotate=45]{\color{black}\footnotesize $i$ (60~pW)};
    \draw (0.446,0.431) node[ rotate=-45]{\color{black}\footnotesize $p$+$s$};
    \draw (0.518,0.982) node[]{\color{black}\footnotesize Grating spectrometer};
    \draw (0.715,0.98) node[]{\color{black}\footnotesize (a)};
    \draw (0.715,0.51) node[]{\color{black}\footnotesize (b)};
 \end{scope}
\end{tikzpicture}
\caption{Nonlinear interferometer for sQFTIR-based hyperspectral mapping. A 660~nm pump laser is focused by an achromatic (AC) lens and triggers spontaneous parametric down-conversion (SPDC) in the nonlinear crystal. The pump, signal, and idler beams are collimated by an off-axis parabolic mirror (OAPM). A dichroic mirror (DM) separates the idler beam from the pump and signal beams, and the idler photons are focused onto the sample using an AC doublet. After reflection, the pump, signal, and idler beams are overlapped in the nonlinear crystal, where SPDC can occur a second time. The resulting signal-photon interference is coupled to the spectrometer via a cold mirror (CM). Insets (a) and (b) display the spatial performance of the system, evaluated using the edge-spread response and the derived point-spread function, as well as an image of groups 4 and 5 of a USAF 1951 resolution test target.}
\label{fig:system}
\end{figure*}

Equations~\ref{eq:SNR} and \ref{eq:SNR_spectroscopy} determine the sensitivity of the detection in both domains in the shot-noise-limited regime. However, in order to express the sensitivity advantage, the total measurement time $T_{\mathrm{acq}}$ must be considered.

In the TD acquisition, the interferogram is measured sequentially, and the available measurement time has to be distributed over the delay positions.
In the FD case, all spectral and depth channels are measured simultaneously during the full exposure time of the line detector, and thus the signals add coherently, while the shot-noise contributions are white and statistically uncorrelated and therefore add incoherently. 

Consequently, the effective time per sample in the TD can be expressed as
\begin{equation}
\tau_{\mathrm{TD}} = \frac{T_{\mathrm{acq}}}{N},
\qquad
B \approx \frac{1}{2\tau_{\mathrm{TD}}}
= \frac{N}{2T_{\mathrm{acq}}},
\end{equation}
whereas in the FD case each spectral channel is integrated over the full acquisition time,
\begin{equation}
\tau = T_{\mathrm{acq}}.
\end{equation}
Therefore, substituting into Eq.~\eqref{eq:SNR}, it can be shown that for the same total acquisition time~\cite{Leitgeb2003, DeBoer2017},
\begin{equation}
\mathrm{SNR}_{\mathrm{FD}}
\approx
\frac{N}{2}\mathrm{SNR}_{\mathrm{TD}}.
\label{eq:SNR_advantage}
\end{equation}

The constant 1/2~\cite{Leitgeb2003} accounts for differences between energy-based (FD) and power-based (TD) measurements due to underlying connections to Nyquist's theorem. This shows that shot-noise-limited FD measurements using line cameras with $N$ pixels deliver a fundamental sensitivity advantage over TD measurements (with a single pixel) that can be viewed as a "multipixel" advantage. In OCT imaging, it is essentially a multiplexing advantage in the sense that all depths are measured simultaneously, the back-reflected waves are coherently summed up, and the depth profile is finally recovered by a Fourier transform. It has to be emphasized again that the SNR expressions are defined in terms of power ratios. For intensity (amplitude) measurements the noise amplitude scales with $\sqrt{N}$ while the signal amplitude scales with $N$ (see Eq.~\ref{eq:SNR_spectroscopy}). Therefore, the net scaling of amplitude-SNR is proportional to $\sqrt{N}$.

\section{Experimental setup of sQFTIR mapping and characterization}\label{sec:exp}
\begin{figure*}[htb]
\centering
    \subfloat[\centering TD comparison of sQFTIR and QFTIR]{{\includegraphics[width=.372\textwidth]{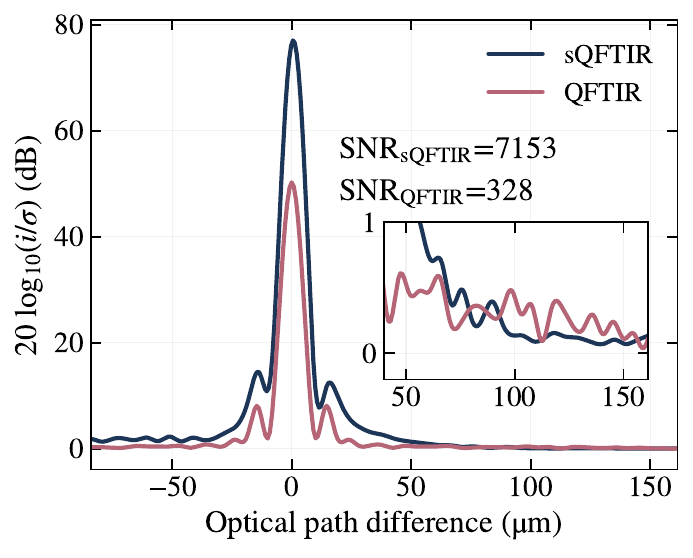} }}%
    \quad
    \subfloat[\centering FD comparison of sQFTIR and QFTIR]{{\includegraphics[width=.372\textwidth]{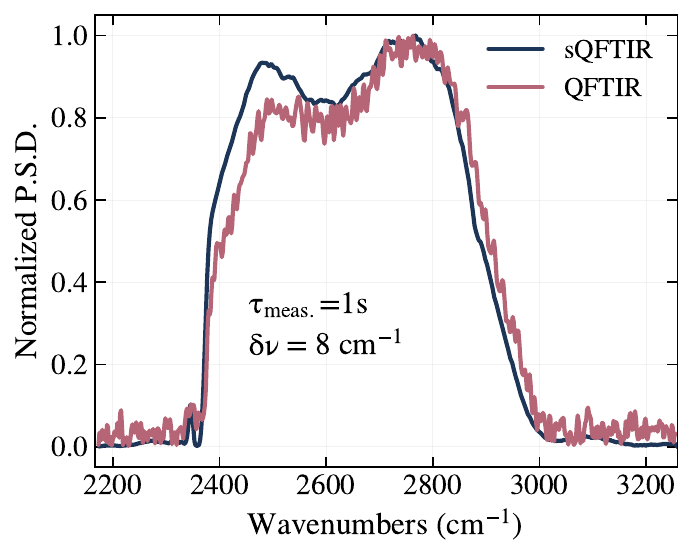} }}%
    \caption{Demonstration of sensitivity advantage: signal-to-noise ratios in the time and Fourier domains for a measurement time of 1~s obtained in the same nonlinear interferometer. (a) Time-domain representation: signal-to-noise ratios ($i/\sigma$) evaluated for sQFTIR and QFTIR from their respective signal magnitudes.
    (b) Frequency-domain representation: sQFTIR and QFTIR spectra calculated from the respective time-domain coherence functions. Differences in the shapes of sQFTIR and QFTIR spectra can be attributed to non-uniform instrument responses. The higher signal-to-noise ratio is preserved between the domains as manifested by Parseval's theorem.}%
    \label{fig:SNR}%
\end{figure*}
For the experimental setup, a nonlinear interferometer with Michelson topology (SU(1,1)), represented in Fig.~\ref{fig:system}, was used. A beam from a pump laser (660~nm, 500~mW, Cobolt Flamenco), magnified by a telescope to a diameter of 2.3~mm,  enters the interferometer via a cold mirror (CM) and is then focused via an achromatic lens (AC, $f=75$~mm) onto a periodically poled KTP (ppKTP) crystal ($l=2.55$~mm, $\Lambda=2.45$~\textmu m, Raicol), thus achieving a focusing parameter of 1.42 (ratio of the crystal length to the confocal parameter). There is a small probability (on the scale of $10^{-9}$) that the pump photons are converted  via SPDC into signal and idler photons in the ppKTP crystal. The SPDC is phase-matched (collinear type-0) for broadband emission~\cite{Vanselow:19}. The pump and emerging signal and idler beams are collimated by an off-axis parabolic mirror (OAPM) and are then directed onto a dichroic mirror (DM) that separates the pump and signal photons from the idler photons. In the combined pump and signal arm, the photons are reflected by a fixed gold mirror back toward the DM. In the idler arm, the photons are focused via a mid-IR AC doublet onto the sample and the back-scattered idler photons are transmitted through the DM towards the nonlinear crystal. Pump, signal, and idler photons pass through the ppKTP crystal a second time, where pump photons could again be converted into signal and idler photons. Due to the operation in the low-gain regime, difference frequency generation between the signal and pump photons can be neglected.

Quantum interference between signal (and also idler) photons due to path indistinguishability occurs in the second pass through the crystal, and the interfering signal photons are transmitted through the CM, coupled into a single-mode fiber, and detected via a grating spectrometer. 
High-resolution and high-speed measurements were performed using an uncooled customized HR6 spectrometer (Ocean Optics, 2048 pixels) and a cooled customized QEPro spectrometer (Ocean Optics, 1024 pixels), respectively.

The imaging capabilities of the system were assessed by mapping a 1951 USAF test target, shown in the insets of Fig.~\ref{fig:system}. The 230 $\times$ 230 pixel map was recorded with a step size of 5~\textmu m and a spectrometer integration time of 200~ms. Intensity values in Fig.~\ref{fig:system}~(b) represent the integrated reflection of the resolution elements. Groups 4 and 5 were identified as suitable for assessing the system resolution, with a corresponding measured idler beam diameter of 12.5~mm (knife-edge method). The square in group 4 was selected to evaluate the resolution limit. Therefore, the edge-spread (ESF) was fitted with a Fermi-function (shown in Fig.~\ref{fig:system}~(a)). The point-spread function (PSF) was then derived~\cite{Marchand:64}, and the resulting resolution limit was determined from the full width at half maximum (FWHM), resulting in 12.3~\textmu m. This was verified by raster scanning the test target, where the 3\textsuperscript{rd} element of group 5 (line-width of 12.4~\textmu m) can still be resolved.

During the characterization stage, TD and FD sampling approaches were directly compared in both domains for the same interferometric arrangement to demonstrate the intrinsic sensitivity advantage.
The determination of the SNR of the sQFTIR modality was performed in an empty interferometer, with the AC doublet removed and the sample replaced by a gold mirror. The integration time of the spectrometer was set to 500~ms and two spectra were averaged to match the standard QFTIR measurement time of 1~s. The OCT-post-processing part of the formalism introduced in Fig.~\ref{fig:flow-diagram} (i.e., subtraction of reference, remapping to $k$-space, inverse Fourier transform, and re-centering) was applied to a FD spectral interferogram and to an FD spectrum in which the idler arm was blocked, i.e., in the absence of interference. The phase offset in the first measurement, shown in Fig.~\ref{fig:SNR}~(a), was set so that the spectral resolution was $20$~cm$^{-1}$. Furthermore, the inherent unbalanced group velocity dispersion~\cite{Zorin2026} was corrected numerically, yielding a symmetric center burst. 
The maximum of the magnitude of the resulting center burst was used to determine $i$ and the standard deviation of the magnitude of the secondary measurement without interference (and hence no center burst; noise floor) was used to determine $\sigma$. 
An SNR of 7153 was calculated according to the spectroscopy convention ($i/\sigma$) and an $\mathrm{SNR}_{\mathrm{FD}}$ of 77.1~dB was further computed by $\mathrm{SNR_{\mathrm{dB}}} = 20\ \mathrm{log_{10}(i/\sigma)}$ (according to OCT convention). The resulting magnitude spectrum, expressed in terms of $\mathrm{SNR_{\mathrm{dB}}}$, is depicted in Fig.~\ref{fig:SNR}~(a). It should be noted that the theoretical shot-noise-limited SNR defined by equation~\ref{eq:SNR} is hard to assess because $\eta$ is a compound function of the instrument and accounts for losses such as coupling and reflection losses, diffraction and quantum efficiency. An $\eta \approx 0.24 $ was estimated under the assumption of shot-noise-dominated performance. 

The same procedure was carried out for the QFTIR modality with a measurement time of 1~s (measurement in the TD via scanning of the combined signal and pump arm) and a spectral resolution of $20$~cm$^{-1}$. In order to correct the unbalanced group velocity dispersion, a Fourier transform was applied to the interferogram. DC subtraction was not necessary, as the detector was operated in the AC mode. 
The SNR was evaluated to be 328 (50.3~dB) using the same formalism as previously described. The magnitudes of the coherence functions, expressed in terms of $\mathrm{SNR_{\mathrm{dB}}}$, are plotted in Fig.~\ref{fig:SNR}~(a). The inset in Fig.~\ref{fig:SNR}~(a) compares the noise floors of the two different measurement modalities. This result is in good agreement with the sensitivity advantage expressed in equation~\ref{eq:SNR_advantage}.  The intensity-based SNR of the FD domain approach should scale with $\sqrt{N/2}$. For the TD intensity SNR of $328$, this would result in an FD intensity $\mathrm{SNR_{FD}}$ of 7494. The 5\% discrepancy between this value and the measured value of 7153 can be attributed to other minor noise contributions (see Eq.~\ref{eq:fullSNR}) and drifts of the signal.

Figure~\ref{fig:SNR}~(b) shows the FD representation of sQFTIR and QFTIR, i.e., reconstructed spectra captured at equalized spectral resolution and measurement time. It can be clearly seen that the sensitivity advantage that was demonstrated for the TD is preserved upon the switch to the FD. For these measurements, a spectral resolution of 8~cm$^{-1}$ and a total measurement time of 1~s were selected for both modalities. The differences in shape of the PSD can be attributed to non-uniform instrument responses, e.g., imperfect alignment of the spectrometer.

\section{Results: Hyperspectral imaging and microspectroscopy} 

\begin{figure}[b!]
\centering
\captionsetup[subfigure]{oneside,margin={0cm,0cm}}
    \subfloat[\centering Hyperspectral image of multilayer foil 2850--2891~cm\textsuperscript{-1}]{\includegraphics[width=.17\textwidth]{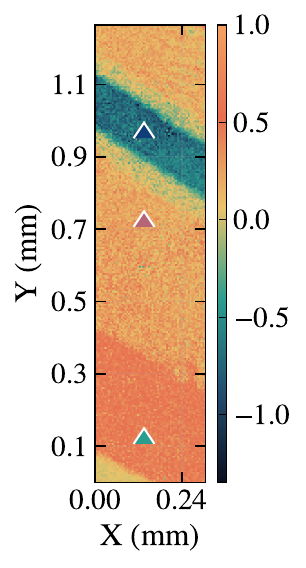} }%
    \quad
    \subfloat[\centering Hyperspectral image of multilayer foil 2836--2860~cm\textsuperscript{-1}]{\includegraphics[width=.17\textwidth]{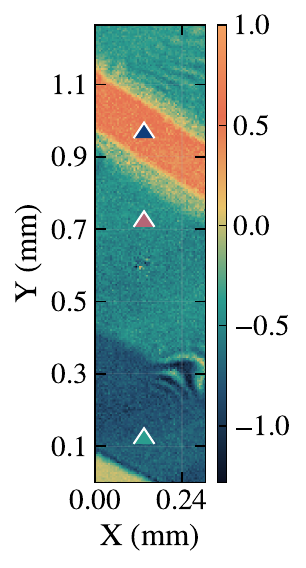} }%
    \quad
    \subfloat[\centering Single absorbance spectra (10~ms integration time) from the hyperspectral image; integration bands used for hyperspectral images are indicated]{\hspace{-2em} \includegraphics[width=.350\textwidth]{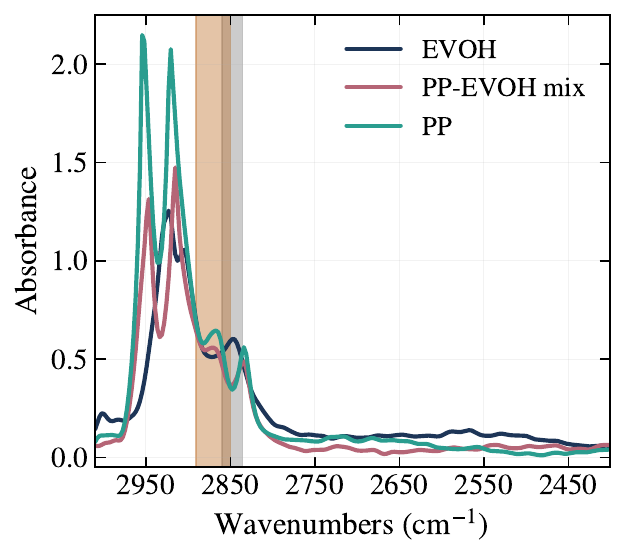} }%
    \caption{Rapid hyperspectral transflection mapping of the cross section (sectioned to a thickness of 5~\textmu m) of a polymer multilayer foil (packaging) with layers of EVOH, PP and PP-EVOH mix recorded at 10~ms per pixel. The positions from which individual (single-pixel) PP, EVOH, and mixed spectra were extracted are indicated with color-coded triangles. Chemical images obtained by integrating absorbance from baseline over (a) 2850--2891~cm\textsuperscript{-1}; (b)  2836--2860~cm\textsuperscript{-1} of the same polymer multilayer foil; (c) absorbance spectra of EVOH, PP, and PP-EVOH mix extracted from the hyperspectral scan. The spectral bands are clearly resolved; the integration bands used to produce the images (a) and (b) are indicated.}%
    \label{fig:multi}%
\end{figure}

In order to evaluate the combined spectral and spatial performance in a real applied scenario, hyperspectral mapping of a cross section of a multilayer polymer packaging film (cut to 5~\textmu m thickness) attached to a gold mirror was performed and the results are depicted in Fig.~\ref{fig:multi}.
The sample is composed of ethylene vinyl alcohol (EVOH), polypropylene (PP), and PP-mix layers. The mix region refers to the recycled material stream from the remaining multilayer foil after the product shapes have been punched out. This foil residue is ground and reintroduced into the process. It may contain varying proportions of PP and EVOH, so it is treated as a mixed multilayer recyclate. The 253~$\times$~61 pixel hyperspectral map was recorded using an integration time of 10~ms per pixel.

The spectral resolution, which is defined as the inverse of the phase offset, $1/\Delta z$, was set to 23.8~cm$^{-1}$. Absorbance values were calculated using a gold reference spectrum obtained from the same scan (lower left corner in the hyperspectral images).
\begin{figure*}[t]
\centering
\captionsetup[subfigure]{oneside,margin={0cm,0cm}}

\begin{minipage}[t]{\textwidth}
\centering

\makebox[\textwidth][c]{%
    \begin{minipage}[t]{0.9\textwidth}
    \raggedleft
    \subfloat[\centering Spectral image of microplastic particle]{%
        \includegraphics[height=.32\textwidth]{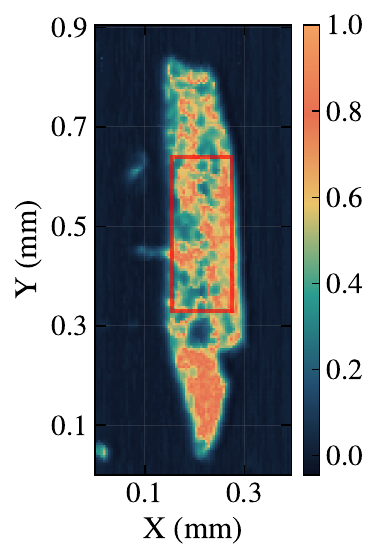}
    }%
    \end{minipage}%
    \hspace{0.08\textwidth}%
    \begin{minipage}[t]{0.9\textwidth}
    \raggedright
    \subfloat[\centering Absorbance spectrum (averaged)]{%
        \includegraphics[height=.32\textwidth]{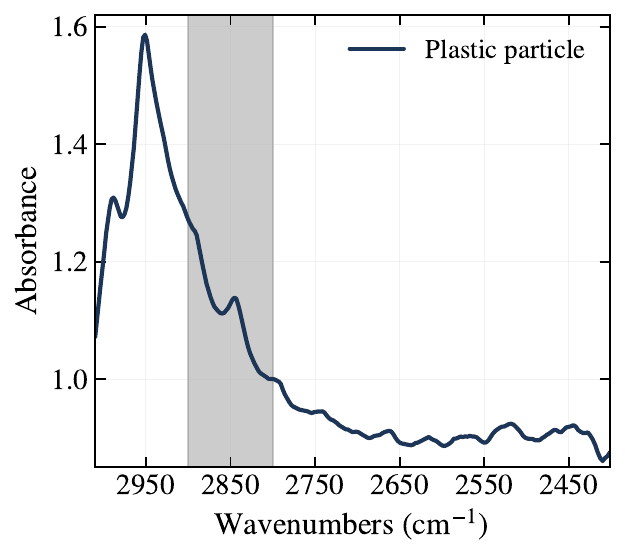}
    }%
    \end{minipage}%
}

\caption{Hyperspectral imaging of a polymer particle in diffuse reflection mode with an integration time of 200~ms per pixel; the area used for spectral averaging is marked with a red rectangle. (a) normalized integrated spectral absorbance image (integrated over 2540--2915~cm\textsuperscript{-1}) of a polymer particle. (b) An area-averaged absorbance spectrum extracted from the hyperspectral scan. A spectrum acquired from a gold mirror was used as the reference. The spectral bands are clearly resolved and the integration band used to produce the image in (a) is indicated in gray.}
\label{fig:particle}

\vspace{1em}
\end{minipage}

\begin{minipage}[t]{\textwidth}
\centering

\makebox[\textwidth][c]{%
    \begin{minipage}[t]{0.9\textwidth}
    \raggedleft
    \subfloat[\centering Spectral images of colon tissue]{%
        \includegraphics[height=.33\textwidth]{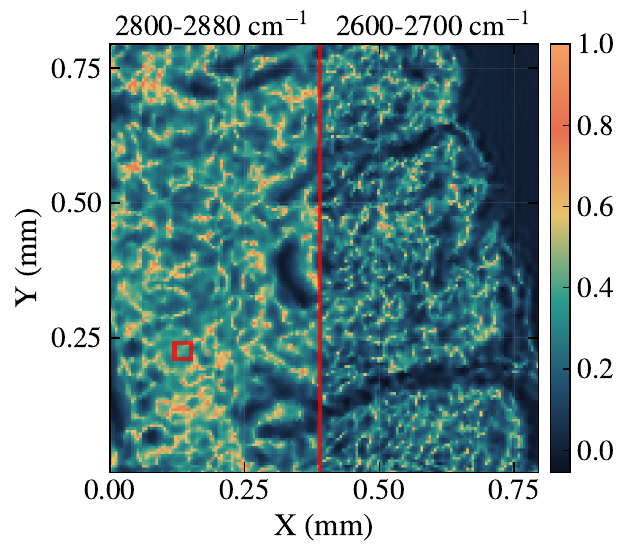}
    }%
    \end{minipage}%
    \hspace{0.08\textwidth}%
    \begin{minipage}[t]{0.9\textwidth}
    \raggedright
    \subfloat[\centering Absorbance spectrum (averaged)]{%
        \includegraphics[height=.32\textwidth]{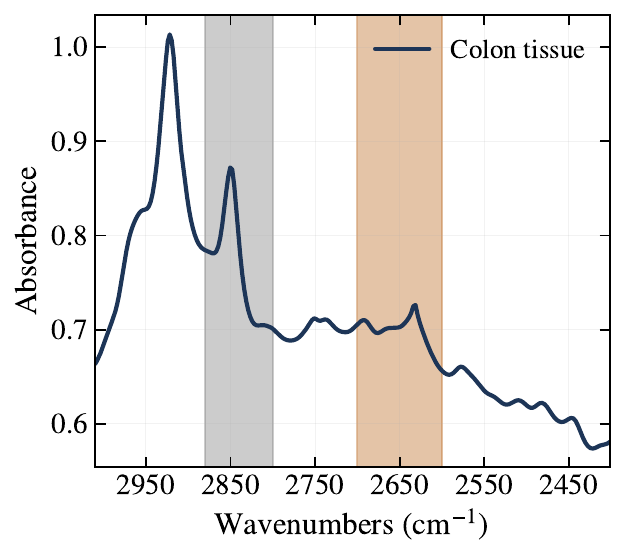}
    }%
    \end{minipage}%
}

\caption{Hyperspectral image of colon tissue recorded in transflection mode: (a) integrated spectral absorbance image (integrated over 2800--2880~cm\textsuperscript{-1} and 2600--2700~cm\textsuperscript{-1}, normalized) of a colon tissue cross section sectioned to a thickness of 5~\textmu m; the sample was applied onto a gold mirror; (b) representative averaged absorbance spectrum (averaging area indicated by a red square in (a)) with an integration time of 200~ms per pixel.}
\label{fig:tissue}

\end{minipage}

\end{figure*}
The integrated absorbance from the baseline of the band from 2850--2891~cm$^{-1}$ is shown in Fig.~\ref{fig:multi}~(a). A second map showing integrated absorbance from the baseline of the band from 2836--2860~cm$^{-1}$ is shown in Fig.~\ref{fig:multi}~(b). Due to the low sample thickness and its flatness, high-quality transflection spectra at the single-pixel level were obtained. Three extracted single-shot spectra of EVOH, PP–EVOH mix and PP, respectively, are shown in Fig.~\ref{fig:multi}~(c) with the integration band used to produce the hyperspectral map indicated.  

In order to further illustrate the capabilities of the sQFTIR-based hyperspectral mapping modality, samples with more complex geometries and strong scattering were investigated. A large microplastic particle (PP), fixed via adhesion to a gold mirror, was mapped with an integration time of 200~ms per pixel and a spectral resolution of 44~cm$^{-1}$. The full 240 $\times$ 80 pixels image is shown in Fig.~\ref{fig:particle}~(a). Each pixel represents the normalized integrated absorbance over the band 2540--2915~cm$^{-1}$. A representative full mid-IR spectrum (averaged over a defined area, indicated by a red rectangle) is shown in Fig.~\ref{fig:particle}~(b). The hyperspectral image displays severe scattering effects common in reflection spectroscopy geometries for samples with a complex and irregular structure. The main indicator of such effects is the large absorbance offset in the spectral region from 2450--2750~cm$^{-1}$. This apparent absorbance can be attributed to scattering since there are no pronounced absorption bands. Additionally, the real absorbance bands of the polymer particle are widened and shifted, which can also be attributed to scattering.

In order to demonstrate the capabilities of the proposed method for biomedical applications, hyperspectral measurements of human colon tissue were performed. The human biological material (reference number: DiDa) was provided by the Institute of Pathology, Paracelsus Medical University (PMU), Salzburg, Austria. The laboratory is certified according to ISO 9001:2015 and operates in accordance with the accreditation standard for medical laboratories ISO 15189:2022~\cite{iso201215189}. The colon sample was sectioned to a thickness of 5~\textmu m, mounted on a gold-coated microscope slide and deparaffinized. The sample thickness and reflective substrate enabled mid-infrared transflection measurements while minimizing spectral interference from the embedding medium and conventional glass substrates. The 160 $\times$ 160 pixel hyperspectral image was acquired with a step size of 5~\textmu m, an integration time of 200~ms per pixel, and a spectral resolution of 44~cm$^{-1}$. The integrated absorbance maps are shown in Fig.~\ref{fig:tissue}~(a). The mid-IR spectral ranges used for integration were 2800~cm$^{-1}$ to 2880~cm$^{-1}$ and 2600~cm$^{-1}$ to 2700~cm$^{-1}$. An averaged absorbance spectrum from the area indicated by the red rectangle is shown in Fig.~\ref{fig:tissue}~(b). Biological samples are highly complex, and mid-IR reflection spectra can exhibit scattering artifacts, as reflected by the offset in the averaged absorbance spectrum. We suppose the derived results display the spatial distributions of lipids, membrane material, protein side chains, and general organic tissue density associated with C-H stretching vibrations.

FTIR hyperspectral imaging is a powerful tool that can chemically map the differences between normal and malignant colonic tissues~\cite{ramesh_ftir_2001,argov_diagnostic_2002,li_vivo_2005}, since the majority of cancer forms manifest themselves in different optical signatures. However, classical FTIR microspectroscopic mapping is expensive and usually requires 1 to 2~s to acquire a single pixel spectrum due to the trade-off between spatial resolution and sensitivity typical for classical instruments based on black-body emitters and low-efficiency mid-IR detection. In this scenario, the proposed sQFTIR method represents a particularly promising alternative.  
\vfill

\section{Conclusion and Outlook}

In this paper, we have introduced and demonstrated a scanless quantum Fourier-transform infrared spectroscopy (sQFTIR) modality based on frequency-domain sensing with undetected photons and a tailored reconstruction protocol. The sQFTIR modality was enabled by unique features inherent to interferometric sensing with undetected photons; the method eliminates optical delay scanning while preserving established and reliable OCT and FTIR post-processing to reconstruct mid-IR spectra. 
Furthermore, we theoretically and experimentally demonstrated that, for comparable acquisition times, frequency-domain acquisition offers an intrinsic sensitivity advantage over time-domain approaches. This advantage arises from spectral multiplexing and coherent summation and amounts to 26.8~dB (a factor of 21.8 according to the spectroscopic definition) in our implementation. 
By harnessing quantum entanglement and Fourier-domain detection, the presented approach enables robust and rapid mid-infrared FTIR spectroscopy using near-infrared detection. 

In the experimental part, a system based on a low-gain nonlinear interferometer and a non-degenerate source of correlated photons was demonstrated. The high speed and sensitivity of the sQFTIR implementation enabled microscopic hyperspectral mapping with high-quality single-pixel spectra acquired in acquisition times as short as 10~ms at a probing power of only 60~pW using a cost-effective all-near-IR detection scheme. This unique combination of capabilities cannot be achieved with state-of-the-art FTIR systems (about 500--1000 times slower), quantum-cascade-laser (QCL)-based mid-IR mapping systems (about 100--200 times slower and requiring higher probe powers) and QCL-based imaging systems (requiring higher probe powers). 
The particular relevance for future biomedical and clinical applications~\cite{Ferguson2025} has been demonstrated in hyperspectral mapping with near-diffraction-limited spatial resolution. 
The performance was verified using complex samples relevant to materials-science and biomedical applications, including multilayer polymer foils, microplastics, and human tissue.

The sQFTIR method is based on frequency-domain measurements and the reconstruction of time-domain coherence functions, analogous to the time-domain interferograms used in classical and quantum FTIR. While this approach provides a fundamental sensitivity advantage, it also introduces a fundamental trade-off that is absent in direct time-domain measurements. This trade-off arises from signal roll-off: larger relative delays are encoded by higher fringe frequencies, which, beyond a certain frequency, cannot be sampled or unambiguously resolved because of the finite spectral resolution of the spectrometer and the Nyquist limit. As in conventional FTIR, the spectral resolution is inversely proportional to the group delay. In sQFTIR, however, the maximum measurable delay is governed by the sampling rate and resolving power of the spectrometer. Achieving high spectral resolution therefore typically requires a custom spectrometer with sufficiently fine spectral sampling. This requirement is nevertheless easier to meet in the near-IR domain, where linear arrays with up to 16,000 pixels are available.

We anticipate that this quantum-based approach has the potential to reshape the landscape of mid-IR spectroscopy due to its robust architecture with no moving parts, high sensitivity, and acquisition speed. Its compatibility with compact and potentially monolithic implementations, together with the elimination of direct mid-IR sources and detectors, may enable cost-efficient in-line mid-IR sensing and biomedical research, e.g., in micro- and nanoplastic studies~\cite{duswald_detection_2025,aichinger_evaluating_2026}, especially when the accessible wavelength regimes are pushed further towards the fingerprint regime~\cite{Schunemann2016,Schunemann2021}.

\section*{Funding}
\noindent This project was co-financed by research subsidies granted by the Government of Upper Austria under the QUICK (Wi-2022-597365/18-Au) and QUANTAN (Wi-2026-67023/3-FrJ) projects. The authors acknowledge funding and support from Österreichische Forschungsförderungsgesellschaft (FFG) under QMIRACT Project (929209) 

\section*{Acknowledgment}
\noindent The authors would like to thank Sven Ramelow and his group members for helpful discussions and lending us a custom dichroic mirror. The authors are grateful to Gregor Langer, Robert Zimmerleiter, and Chiara Lindner for fruitful discussions.

\noindent The authors express their gratitude to Karl Sotlar and Johannes Heigl from the Institute of Pathology, Paracelsus Medical University (PMU), Salzburg, Austria for providing human tissue samples.

\section*{Disclosures}

\noindent The authors declare no conflicts of interest.

\section*{Data availability} Data underlying the results presented in this paper are not publicly available at this time but may be obtained from the authors upon reasonable request.

\bibliography{main}

\clearpage
\onecolumn
\phantomsection
\label{sec:SUPPL}
\setcounter{section}{0}
\setcounter{figure}{0}
\setcounter{table}{0}
\setcounter{equation}{0}

\renewcommand{\thesection}{S\arabic{section}}
\renewcommand{\thesubsection}{S\arabic{section}.\arabic{subsection}}
\renewcommand{\thefigure}{S\arabic{figure}}
\renewcommand{\thetable}{S\arabic{table}}
\renewcommand{\theequation}{S\arabic{equation}}

\begin{center}
{\Large \textbf{Supplementary document}}\\[1em]
{\large \textbf {Scanless quantum Fourier-transform mid-infrared spectroscopy for rapid high-sensitivity hyperspectral mapping}}\\[0.5em]
Paul Gattinger, Kristina Duswald, Andreas W. Schell, Markus Brandstetter, Bettina Heise, and Ivan Zorin
\end{center}

\begin{onecoltextblock}
This Supplementary Material provides an extended description of the materials and methods, together with additional details on the reconstruction protocol. In particular, it expands description on the individual steps shown in the reconstruction flow diagram and presents the corresponding mathematical formulation.
\end{onecoltextblock}

\vspace{1em}

\section*{Mathematical description of the reconstruction protocol}

In the subsequent mathematical derivations, high-order dispersion terms are neglected. The detected
frequency-domain spectral interferograms recorded in the signal domain can be represented as~\cite{Zorin2026, Vanselow:19}:
\begin{equation}
I(\omega_s)
=
2|f(\omega_s)|^2
\left(
1+r\cos\left[D_s-\omega_s\Delta\tau_g\right]
\right),
\end{equation}
where $f(\omega_s)$ denotes the probability amplitude associated with the
signal mode $\omega_s$, $r$ denotes an effective reflectivity amplitude (for simplicity can be assumed $r=1$),
$\Delta\tau_g$ denotes an effective group-delay mismatch, and $D_s$ denotes
a constant phase term. The term $D_s$ includes the static phase offset and the center-frequency-dependent contribution $\omega_{s0}\Delta\tau_g$. The constant $D_s$ term can be omitted for description of the frequency- to time-domain sQFTIR reconstruction.

After subtraction of the reference spectrum (DC component, an SPDC signal emission recorded by a spectrometer in the absence of interference), the
oscillatory (zero-centered) component can be defined as:
\begin{equation}
I_{\AC}(\omega_s)
=
2|f(\omega_s)|^2
\cos\left[\Delta\tau_g\omega_s\right].
\end{equation}

The energy correlation between idler and signal photons~\cite{BarretoLemos:22}:
\begin{equation}
\omega_p=\omega_s+\omega_i,
\end{equation}
can be used to remap the signal frequencies to the corresponding ($k$-linearized if recorded in wavelengths $\lambda_s$) frequencies of idler photons $\omega_i$.

The remapped interferometric signal can therefore be rewritten for idler photons as:
\begin{equation}
I_{\AC}(\omega_i)
\propto
A(\omega_i)
\cos\left[
\frac{Z_0}{c}\omega_i
\right],
\label{I_ac_idler}
\end{equation}
where $A(\omega_i)$ is the power spectral density of idler photons contributing to bi-photon interference,
$c$ is the speed of light, and $Z_0=c\Delta\tau_g$
is an effective optical path difference associated with the group-delay
mismatch. Due to the round-trip, $Z_0=2z_0$, Eq.~\ref{I_ac_idler} can also be written as:
\begin{equation}
I_{\AC}(\omega_i)
\propto
A(\omega_i)
\cos\left[
\frac{2z_0}{c}\omega_i
\right],
\end{equation}

where $z_0$ is a fixed offset in the nonlinear interferometer.

The spectral interferometric idler pattern defined in Eq.~\ref{I_ac_idler} can be represented in the $k$-space that is Fourier-conjugate to the optical path difference to emphasize the connection:

\begin{equation}
I_{\AC}(k_i)
\propto
A(k_i)\cos(Z_0k_i).
\end{equation}

Using Euler's formula one can derive:
\begin{equation}
\cos(Z_0k_i)
=
\frac{1}{2}e^{\ii Z_0k_i}
+
\frac{1}{2}e^{-\ii Z_0k_i},
\end{equation}
the frequency-domain spectral interferogram can be rewritten as:
\begin{equation}
I_{\AC}(k_i)
=
A(k_i)
\left(
\frac{1}{2}e^{\ii Z_0 k_i}
+
\frac{1}{2}e^{-\ii Z_0 k_i}
\right).
\end{equation}

The time-domain representation of the idler-domain spectral
density $A(k_i)$ can be defined by the inverse Fourier transform as:

\begin{equation}
\Gamma(z)
=
\mathscr{F}^{-1}
\left\{
A(k_i)
\right\}
=
\frac{1}{2\pi}
\int_{-\infty}^{\infty}
A(k_i)\, e^{i k_i z}\, \mathrm{d}k_i .
\end{equation}

The time-domain function $\Gamma(Z)$ is not assumed to have an analytical form.
For instance, for an ideal infinite and flat idler spectral profile it approaches a delta-like
burst, for a Gaussian spectral envelope it has a Gaussian shape, and for a
measured finite spectrum it corresponds to the finite coherence burst determined
by the measured spectral envelope and system response.

The shift of the central burst in the time-spatial domain ($z$-coordinate) from the zero-delay plane can be expressed using the Fourier shift theorem as a multiplication of $A(k_i)$ by a phasor in the frequency $k_i$ domain~\cite{bracewell2000fourier}:

\begin{equation}
\mathscr{F}^{-1}
\left\{
A(k_i)e^{-\ii Z_0 k_i}
\right\}
=
\Gamma(Z-Z_0),
\label{inv_ft_gamma}
\end{equation}

and

\begin{equation}
\mathscr{F}^{-1}
\left\{
A(k_i)e^{\ii Z_0 k_i}
\right\}
=
\Gamma(Z+Z_0).
\end{equation}
Thus, the time-domain representation contains two shifted coherence bursts:
\begin{equation}
I_z(Z)
\propto
\frac{1}{2}\Gamma(Z-Z_0)
+
\frac{1}{2}\Gamma(Z+Z_0).
\end{equation}

In order to implement sQFTIR protocol, a single coherence burst centered at $+Z_0$ is isolated by
multiplication with a window function:
\begin{equation}
I_{z_+}(Z)
=
W_+(Z)\,I_z(Z).
\end{equation}
Here, $W_+(Z)$ denotes a rectangular window centered at the burst position
$Z_0$ so that:
\begin{equation}
W_+(Z)
=
\begin{cases}
1, & |Z-Z_0|\leq \frac{\Delta Z_w}{2},\\[4pt]
0, & |Z-Z_0|> \frac{\Delta Z_w}{2}.
\end{cases}
\end{equation}
Thus, the windowed time-domain coherence function is derived as:
\begin{equation}
I_{z_+}(Z)
\propto
\frac{1}{2}\Gamma(Z-Z_0).
\label{windowed}
\end{equation}

\begin{figure}[ht]
\centering

\subfloat[\centering sQFTIR retrieved TD signals]{%
\begin{minipage}[t]{.45\textwidth}
\vspace{0pt}
\centering
\includegraphics[width=\linewidth]{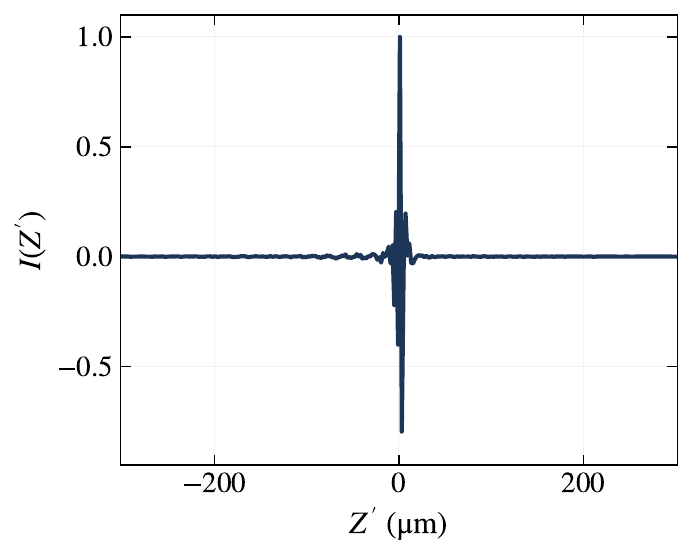}
\end{minipage}%
}
\quad
\subfloat[\centering QFTIR measured TD signals]{%
\begin{minipage}[t]{.445\textwidth}
\vspace{0pt}
\centering
\includegraphics[width=\linewidth]{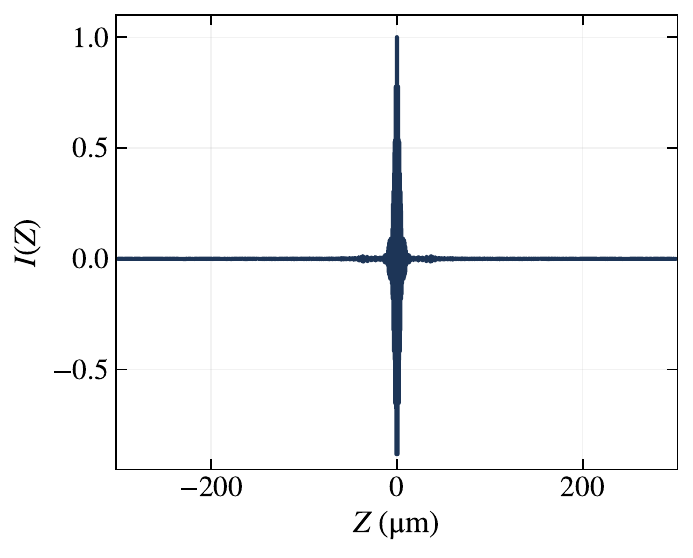}
\end{minipage}%
}

\caption{Time-domain interferograms used for subsequent reconstruction of mid-IR spectra measured in the related Fourier-domain and time-domain modalities, respectively.}
\label{fig:sigs}
\end{figure}

The isolated time-domain coherence function contains the same information as time-domain FTIR and QFTIR interferograms sampled directly using spatial scanning. Consequently, the retrieved signal replicates the time-domain signals obtained directly during the measurement using these methods. See Fig.~\ref{fig:sigs}.

Therefore, the selected spatial range:
\begin{equation}
d \equiv \Delta Z_w=2\cdot\Delta z_w
\end{equation}
where $\Delta z_w$ is a single-sided range, determines, akin to the scanning range in classical scan-based FTIR, the spectral resolution that can be achieved with the sQFTIR method. 
The resolution for a conventional spectroscopic wavenumber can be written as a reciprocal (for the single-sided interferogram):
\begin{equation}
\delta \tilde{\nu}_i \approx \frac{2}{d}.
\end{equation}

After the selection of the burst, an additional apodization function can be applied to $I_{z_+}(Z)$ before the second Fourier transform.
In order to reconstruct smooth sQFTIR interferograms, the selected coherence burst is shifted to a virtual zero-delay plane before the Fourier transform by its original offset of $Z_0=2z_0$: 

\begin{equation}
Z'=Z-Z_0.
\end{equation}

Thus, it is possible to rewrite Eq.~\ref{windowed} for the substituted variable $Z'$, defining the centered coherence function $I_{z_0}(Z')$:
\begin{equation}
I_{z_0}(Z')
\propto
\frac{1}{2}\Gamma(Z').
\end{equation}

In the recentered coordinate $Z'$, the original coherence burst located at $Z=Z_0$ is re-positioned to $Z=0$. 

The coordinate shift $Z'=Z-Z_0$ corresponds to multiplication of the Fourier transform of the time-domain burst Eq.~\ref{windowed} (see original kernel in Eq.~\ref{inv_ft_gamma}) by the inverse
phase factor $e^{\ii Z_0 k_i}$ in the $k_i$ domain:

\begin{equation}
\mathscr{F}
\left\{
I_{z_0}(Z')
\right\}
=
e^{\ii Z_0 k_i}
\mathscr{F}
\left\{
I_{z_+}(Z)
\right\}
\end{equation}

Thus:
\begin{equation}
\mathscr{F}
\left\{
I_{z_0}(Z')
\right\}
=
\frac{1}{2}A(k_i), 
\end{equation}

where $\mathscr{F}$ denotes the Fourier transform to $k_i$ space.

Thus, the recentering operation removes the carrier delay associated with the
spectral fringes while preserving the relative spatial-frequency sampling and
the reconstructed $k_i$ axis. The time-domain central burst is therefore
processed analogously to classical and quantum scan-based FTIR approaches.

Since idler photons remain undetected, the retrieved spectral profile represents the contribution of emitted photons to bi-photon interference, as determined by the visibility $\upnu(k_i)$ and instrument transfer function $H(k_i)$:

\begin{equation}
A(k_i)
=
\upnu(k_i) \cdot H(k_i)\cdot\tilde{S}(k_i),
\end{equation}
where $\tilde{S}(k_i)$ is the total number of emitted photons, i.e., 
\begin{equation}
\tilde{S}(k_i) = |f(\omega_i)|^2.
\end{equation}

In general, $H(k_i)$ accounts for the drop of visibility, e.g., due to the aberrations in the frequency-domain spectrometer and roll-off caused by limited spectral resolution.

The spectrum $A(k_i)$ can be rewritten in angular frequencies as $\tilde{S}(\omega_i)$ using
$\omega_i=c k_i$ or in conventional wavenumbers $A(\tilde{\nu}_i)$ using $\tilde{\nu}_i=k/2\pi$.

$A(k_i)$ represents a quantity proportional to the
idler-domain photon-pair spectral density that contributes to the
nonlinear-interferometric interference term. In other words, it is a share of idler photons that remain indistinguishable
between the first and second passes, i.e., according to the sensing with undetected photons paradigm, it is an idler spectrum inferred from the modulation amplitude of the correlated signal photons.

The same procedure can be applied to both reference and sample measurements to derive IR absorption
or dispersion information using, for instance, the Beer-Lambert law and Kramers–Kronig relations.

\end{document}